\documentclass[twoside]{LCWS11}
\usepackage[latin1]{inputenc}
\usepackage{graphicx,epsfig,color}
\usepackage{wrapfig,rotating}
\usepackage{amssymb,amsmath,array}
\usepackage{cite}
\begin{document}
\title{Mass and Cross Section Measurement of light-flavored Squarks  at CLIC} 

\author{Frank Simon, Lars Weuste
\vspace{.3cm}\\
Max-Planck-Institut f\"ur Physik, M\"unchen, Germany
\vspace{.1cm}\\
Excellence Cluster `Universe', TU M\"unchen,  Garching, Germany
}

\maketitle

\begin{abstract}
We present a study of the prospects for the measurement of TeV-scale
light-flavored right-squark masses and the corresponding production cross section at a
3 TeV $e^+e^-$ collider based on CLIC technology. The analysis,
performed in the framework of the CLIC Conceptual Design Report, is based on
full Geant4 simulations of the CLIC\_ILD detector concept, including standard
model physics background and machine related hadronic background from two-photon
processes. The events are reconstructed using particle flow event
reconstruction, and the mass is obtained from a template
fit built from generator-level simulations with smearing to parametrize the
detector response. For an integrated luminosity of 2 ab$^{-1}$, a
statistical precision of \mbox{5.9 GeV}, corresponding to $0.52\%$, is
obtained for unseparated first and second generation right squarks. For the
combined cross section, a precision of 0.07 fb, corresponding to
$5\%$, is obtained.
\end{abstract}

\section{Introduction}

Future high energy $e^+e^-$ colliders are precision tools for the discovery and the spectroscopy of new particles expected beyond the Standard Model. One attractive extension of the Standard Model is supersymmetry, which predicts a rich spectrum of new particles, one superpartner for each standard model particle. These new particles are expected to have masses in the range from about 100 GeV to a few TeV, and  are thus coming within reach of modern colliders. In the next years, the LHC is expected to provide a decisive answer on the question of the existence of TeV-scale supersymmetry. In typical scenarios, the superpartners of the light quarks ($u, d, s, c$) are among the heaviest superparticles, requiring energies in excess of 1 TeV for pair production. Early LHC results have already placed stringent limits on squark masses, reaching up to about 1 TeV in constrained models \cite{Aad:2011ib, Chatrchyan:2011zy}. The study of these particles at an $e^+e^-$ linear collider requires a multi-TeV machine such as the proposed Compact Linear Collider CLIC \cite{Assmann:2000hg}. 

Here, we consider the production and decay of light-flavored right-squarks in a R-parity conserving SUSY mSUGRA model \cite{LCD-2011-016} where the gluino is heavier than the squarks. In such a scenario, the right-squarks decay essentially exclusively into the lightest neutralino and their standard model partner, resulting in an event signature of two energetic jets and large missing energy,
\begin{equation*}
e^+e^- \, \rightarrow \, \tilde{q_R}\bar{\tilde{q_R}}\,  \rightarrow \, q\bar{q}\chi^0_1\chi^0_1 .
\end{equation*}
The squark masses in the considered model are 1.116 TeV and 1.126 TeV for the first and second generation down-type and up-type squarks, respectively. The combined production cross section, taking into account the CLIC beam energy spectrum at 3 TeV, is 1.47 fb, with a ratio of 4:1 for up- compared to down-type particles. The mass of the lightest neutralino in this scenario is 0.328 TeV.

In the framework of the CLIC Conceptual Design Report \cite{CLIC_CDR}, this process was studied as a benchmark to evaluate the physics performance for generic new physics signatures with high-energy jets, missing energy and the rejection of high-cross-section Standard Model backgrounds. 

\section{Experimental conditions and event reconstruction at CLIC}

The experimental conditions at CLIC, summarized in detail in \cite{CLIC_CDR}, are characterized by high levels of coherent and incoherent $e^+e^-$ pair background as well as mini-jet background originating from $\gamma\gamma \rightarrow$ hadrons processes, due to the high luminosity and high collision energy. Together with the high bunch crossing frequency of 2 GHz,  the latter leads to pile-up of hadronic energy deposits in the detector, in particular in the low-angle regions. This places strict requirements on the event reconstruction and the timing capabilities of all detector components. 

The events are reconstructed using the PandoraPFA particle flow algorithm \cite{Thomson:2009rp}, which determines the timing of each reconstructed particle, allowing the rejection of out-of-time contributions most likely to be background. In addition to the event reconstruction itself, jet finding contributes to the elimination of $\gamma\gamma \rightarrow$ hadrons background. Several jet finders have been evaluated in the course of this study. The best performance was achieved with an exclusive $k_t$ algorithm \cite{Cacciari:2005hq} for hadron colliders, clustering the event into exactly two jets. In this algorithm, the two-particle distance is given by the pseudorapidity $\eta$ and the azimuthal angle $\phi$, resulting in increased distances in the forward region, reducing the sensitivity to background particles. Different jet size parameters, which govern the amount of energy rejected by assigning it to beam jets, were studied, with the best performance observed for $R\, = \, 0.7$.

\section{Squark mass measurement technique}

\begin{wrapfigure}{r}{0.45\columnwidth}
\begin{center}
  \includegraphics[width=.43\columnwidth]{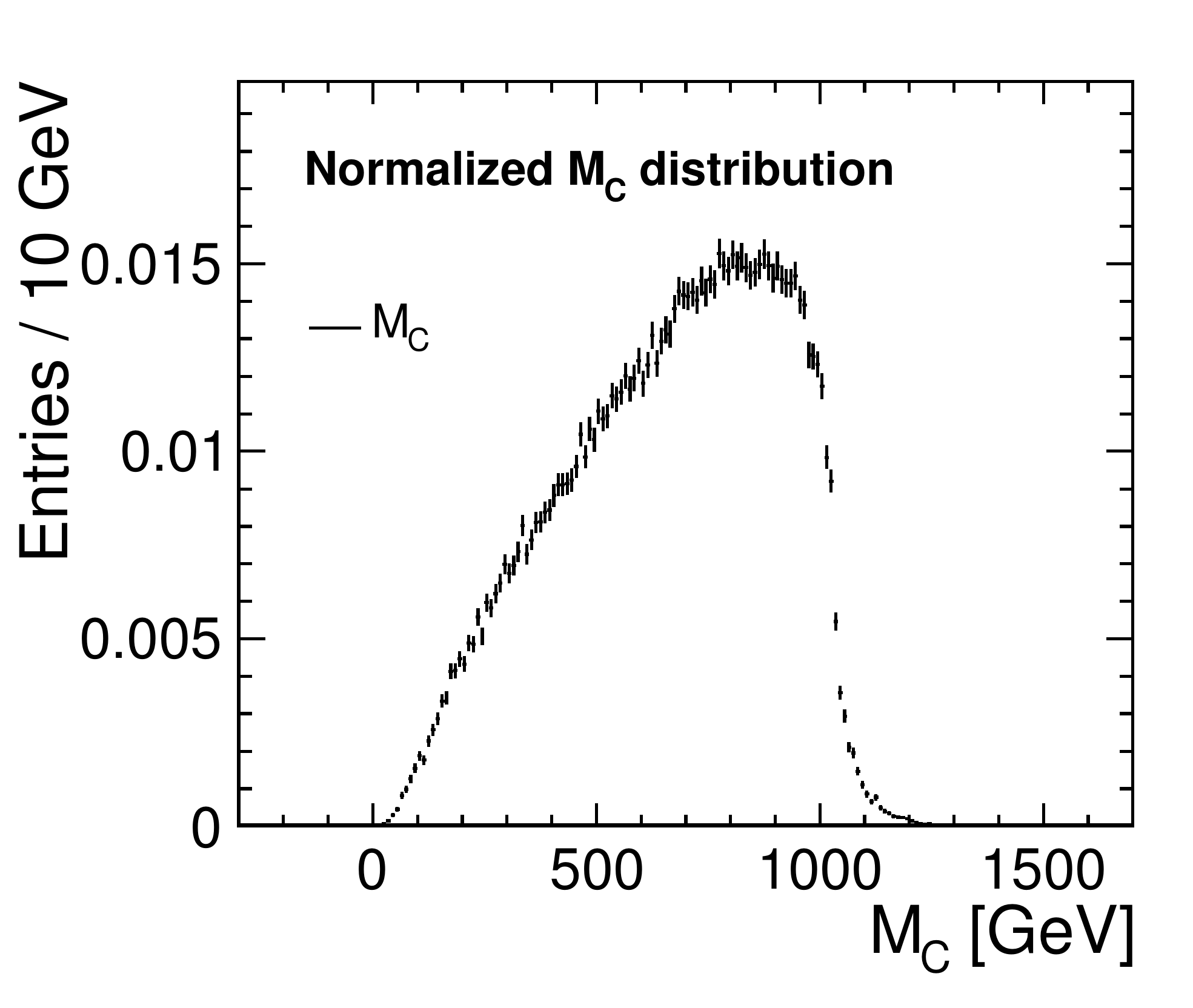}
  \caption{The $M_C$ distribution including effects of the
  beam energy spectrum.}
  \label{fig:Techniques:MC}
\end{center}
\end{wrapfigure}

The distribution of jet energies allows in principle the simultaneous measurement of the squark mass and of the mass of the lightest neutralino from the upper and lower edge of the jet energy distribution. However, this distribution, in particular the lower edge given by low-energetic jets, suffers significantly from standard model background, making precision measurements difficult.  
 
Instead, it is assumed that the mass of the lightest neutralino will be measured with satisfactory precision in processes with higher cross sections and less background sensitivity, such as slepton production and decay \cite{Blaising:2012vd}. With this additional knowledge, the extraction of the squark mass from distributions with a single kinematic edge becomes possible.  

Since the distribution of the center of mass energy at a 3 TeV CLIC collider has a substantial tail towards lower energies due to beamstrahlung, with only 34\% of the luminosity in the top 1\% of the energy, methods which do not rely on the knowledge of the precise center of mass energy are advantageous \cite{LCD-2010-012}. 

One such technique is the variable $M_C$  \cite{Tovey:2008ui}, which uses the momenta of the two observed jets to form a modified invariant mass which is invariant under contra-linear boosts of equal magnitude of the two squarks, and thus independent of the center of mass energy. $M_C$ is given by
\begin{eqnarray}
M_C &=& \sqrt{(E_{q,1} + E_{q,2})^2 - (\vec{p}_{q,1} - \vec{p}_{q,2})^2}\\
&=& \sqrt{2 (E_{1}E_{2} + \vec{p}_{1} \cdot \vec{p}_{2})},
\end{eqnarray}
where $E_{1}$, $\vec{p}_{1}$ and $E_{2}$, $\vec{p}_{2}$ are the energies and
three momenta of the two visible final-state quarks (jets), respectively. The
distribution of $M_C$, taken from generator-level events without background, but with the CLIC beam energy spectrum and with jet finding applied, is shown in Figure \ref{fig:Techniques:MC}. It is
bounded from above by
\begin{equation}
M_C^{max} = \frac{m_{\tilde{q}}^2 - m_\chi^2}{m_{\tilde{q}}},
\end{equation}
providing direct sensitivity to the squark mass.

The construction of $M_C$ assumes that the center of mass system of the
collision is at rest in the detector system, which evidently is not the case at
CLIC due to beamstrahlung and initial state radiation. Still, the boost of the
collision system with respect to the laboratory frame is typically quite small,
making it advantageous to use the complete available information, and not just
transverse observables, as would be done at hadron colliders. The beam energy
spectrum leads to a distortion of the edge of the $M_C$ distribution, as
discussed in detail in  \cite{LCD-2010-012}. The best precision on the squark
mass can thus be obtained from template fits which take the effect of the beam
energy spectrum of CLIC into account.

\section{Simulation and signal selection}

To study the physics performance of CLIC for squark production, signal and background samples have been produced with the WHIZARD event generator \cite{Kilian:2007gr} with fragmentation and hadronization provided by PYTHIA \cite{Sjostrand:2006za}, and have been fully simulated with a detailed GEANT4 \cite{Agostinelli:2002hh} model of the CLIC\_ILD \cite{CLIC_CDR, CLIC_ILD} detector. This detector model is based on the ILD \cite{:2010zzd} detector model for ILC, with some CLIC-specific modifications which account for the higher energy and the different background conditions. Before reconstruction, all events are overlayed with $\gamma\gamma \rightarrow$ hadrons background corresponding to 60 bunch crossings (30 ns). In the event reconstruction, realistic timing cuts are applied inside the particle flow algorithm to reduce the impact of background. Finally, the events are clustered into two jets as discussed above. The present analysis is based on an integrated luminosity of 2 ab$^{-1}$, with additional data sets produced for the training of multivariate analysis techniques discussed below. 

In addition to the reduction of $\gamma\gamma \rightarrow$ hadrons background, the rejection on non-squark physics background is one of the main challenges of
the present analysis. Since the signal signature of two jets and missing energy
is rather generic, high cross section standard model processes contribute to the
background. The study of various background channels has shown that the standard model contributions which are hardest to reject are those with genuine missing energy from neutrinos in the final state. The three major background channels are $\tau\tau\nu\nu$, $qq\nu\nu$ and $qqe\nu$. All three have  cross sections which are 2 to 3 orders of magnitude above the cross section of light-flavor right-squark production.

\begin{figure}[htp]
\begin{center}
  	\includegraphics[width=.48\linewidth]{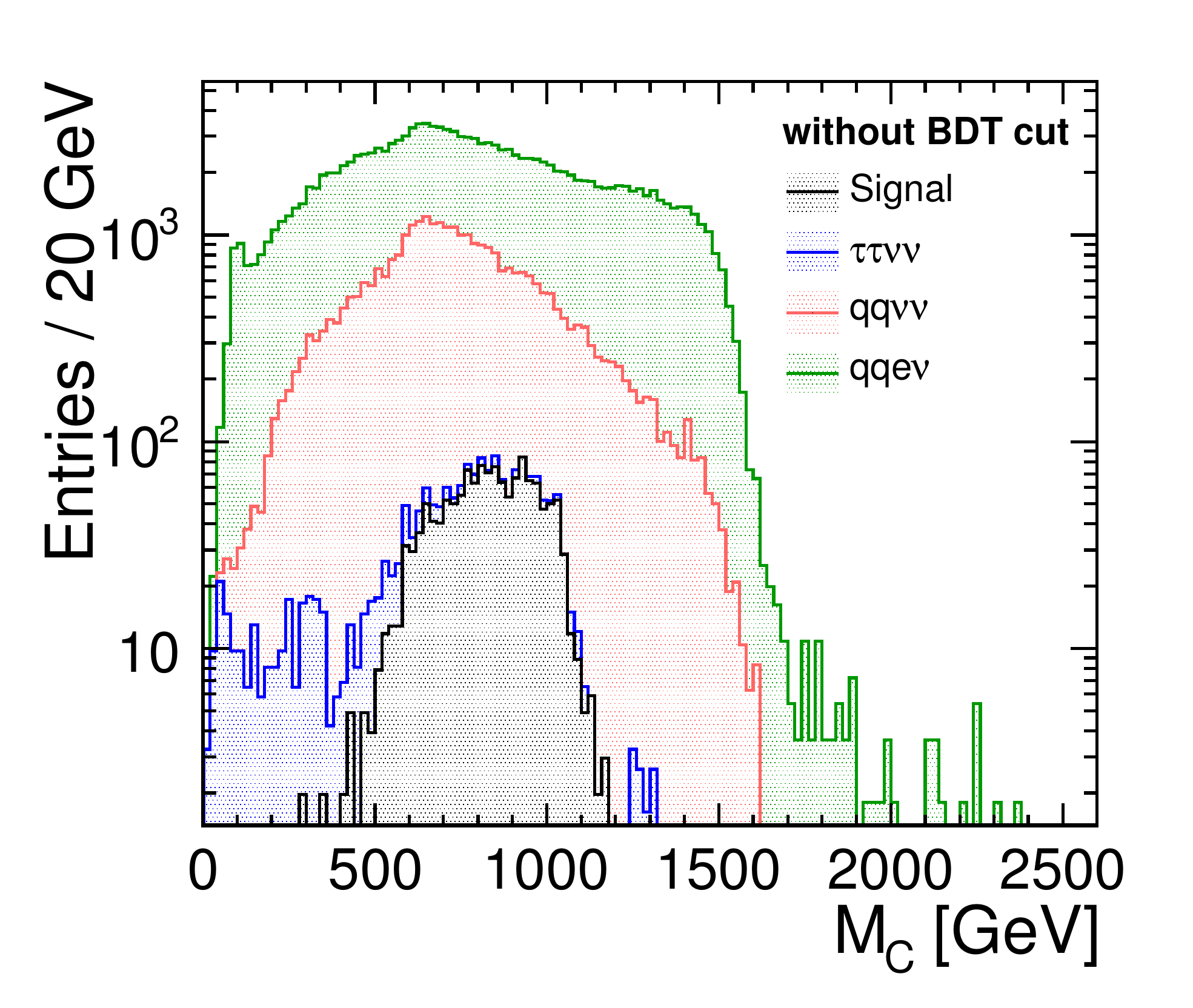}
\hfill
  	\includegraphics[width=.48\linewidth]{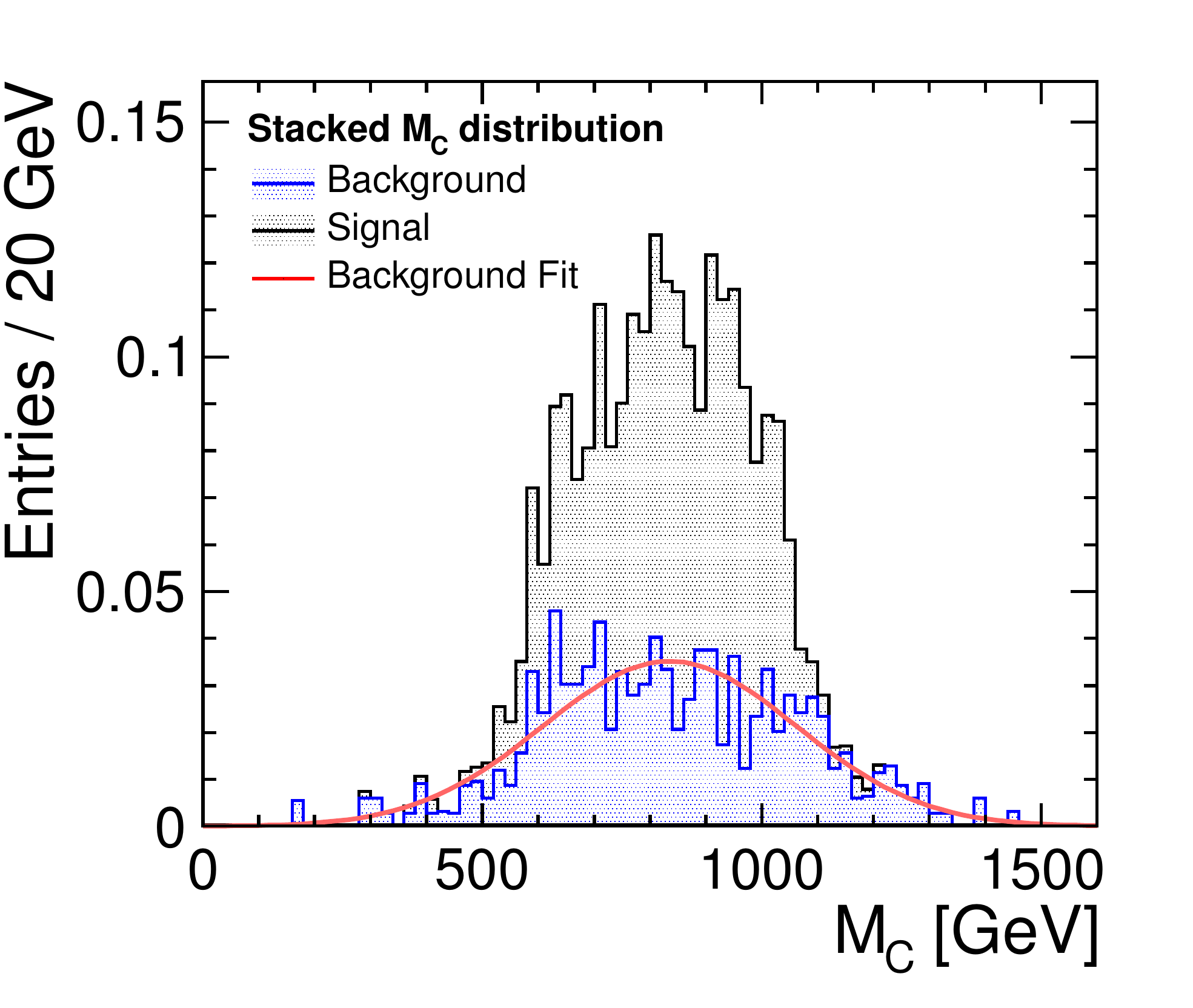}
 
  \caption{{\it Left:} 
  $M_{C}$ distribution stacked with all considered backgrounds after a cut on missing $p_t > 600$ GeV. {\it Right:} Signal and combined background distribution after cut on the output of the Boosted Decision Tree. The fit is used as a parametrization of the background distribution to allow a background subtraction for the mass determination.}
  \label{fig:Background:MC}
\end{center}
\end{figure}

 A high signal purity, in particular in the region of the kinematic edge of the distribution, is crucial to obtain a precise mass measurement. This requires a reduction of the background by more than a factor of 1000. A cut on missing transverse moment can provide quite effective background rejection at the first stage of the analysis. Requiring a measured missing $p_t\, >\, 600$ GeV reduces the dominating background channels to approximately $10^{-2}$, in the case of the $\tau\tau\nu\nu$ final state even to $2\,\times\,10^{-3}$ of their original cross section, while reducing the signal to 0.485. 
 
 This reduction, however, is insufficient to provide a clean separation of signal and background, as shown in Figure \ref{fig:Background:MC} {\it left}, where background processes still dominate the signal by more than an order of magnitude. Further reduction of standard model background is achieved with a boosted decision tree (BDT), as implemented in the TMVA toolkit \cite{Hocker:2007ht}. It provides a separation of signal and background using event shape information, particle multiplicities and energy distributions as well as other discriminating variables. The BDT is trained with signal and background samples not used in the analysis. With the BDT selection, a clean identification of the signal is achieved, demonstrated in Figure  \ref{fig:Background:MC} {\it right}. The overall signal efficiency is 36.1\%, with a signal significance of $S/\sqrt{S+B}\, =\, 25.7$. The background rejection procedure leaves the upper edge of the $M_C$ distribution unchanged, providing the basis for precision mass measurements. The shape in the low $M_C$ region is significantly altered, primarily due to the missing $p_t$ cut, which precludes low $M_C$ values. Since this region of the distribution does not provide sensitivity to the squark mass, this does not affect the measurement. 
 
 For the determination of the squark mass using a template fit, the remaining background is subtracted from the data distribution using a simple parametrization of the background shape determined from a statistically independent background sample.

\section{Results}

\begin{figure}[htb]
\begin{center}
  \includegraphics[width=.49\linewidth]{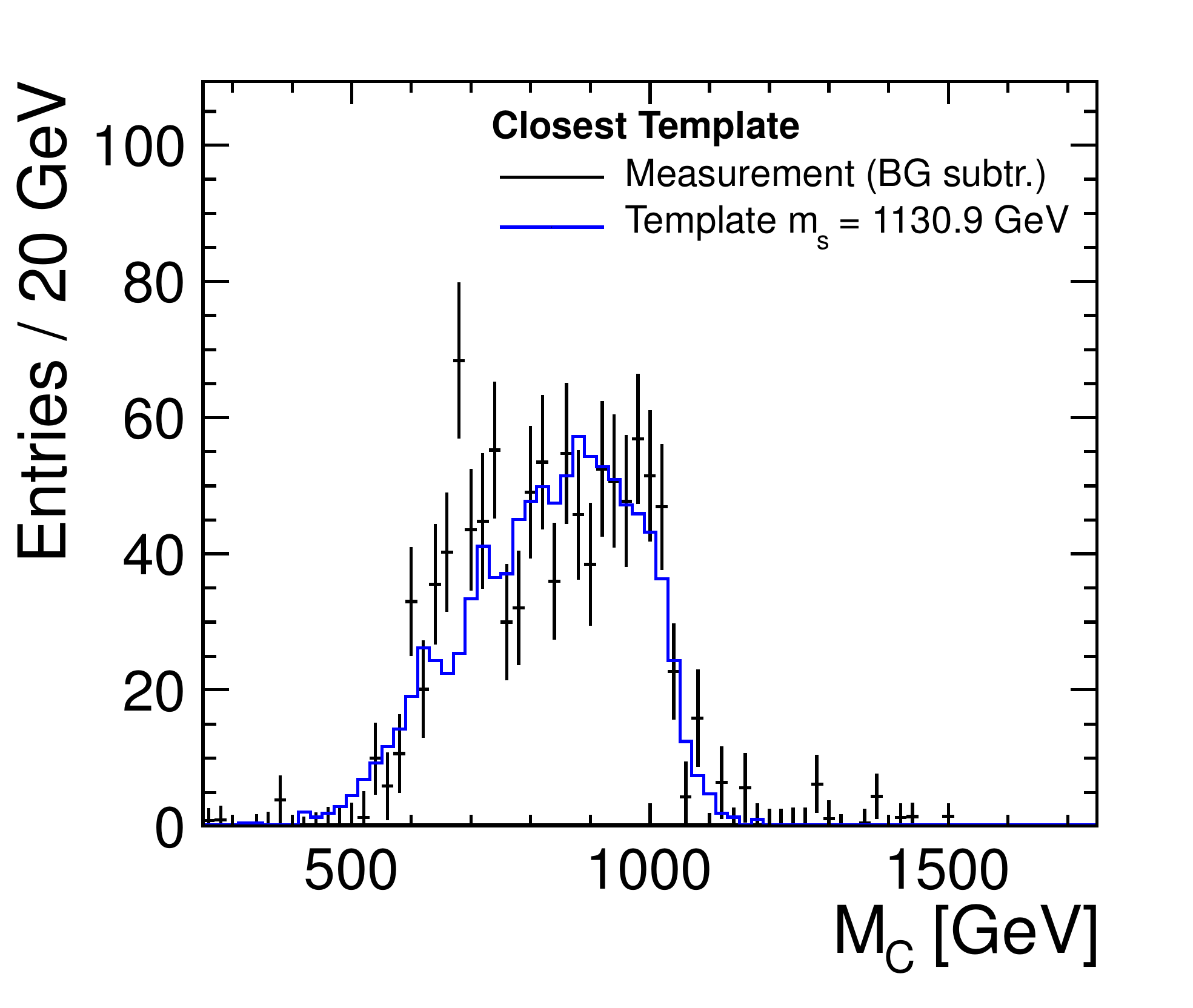}
  \hfill
  \includegraphics[width=.49\linewidth]{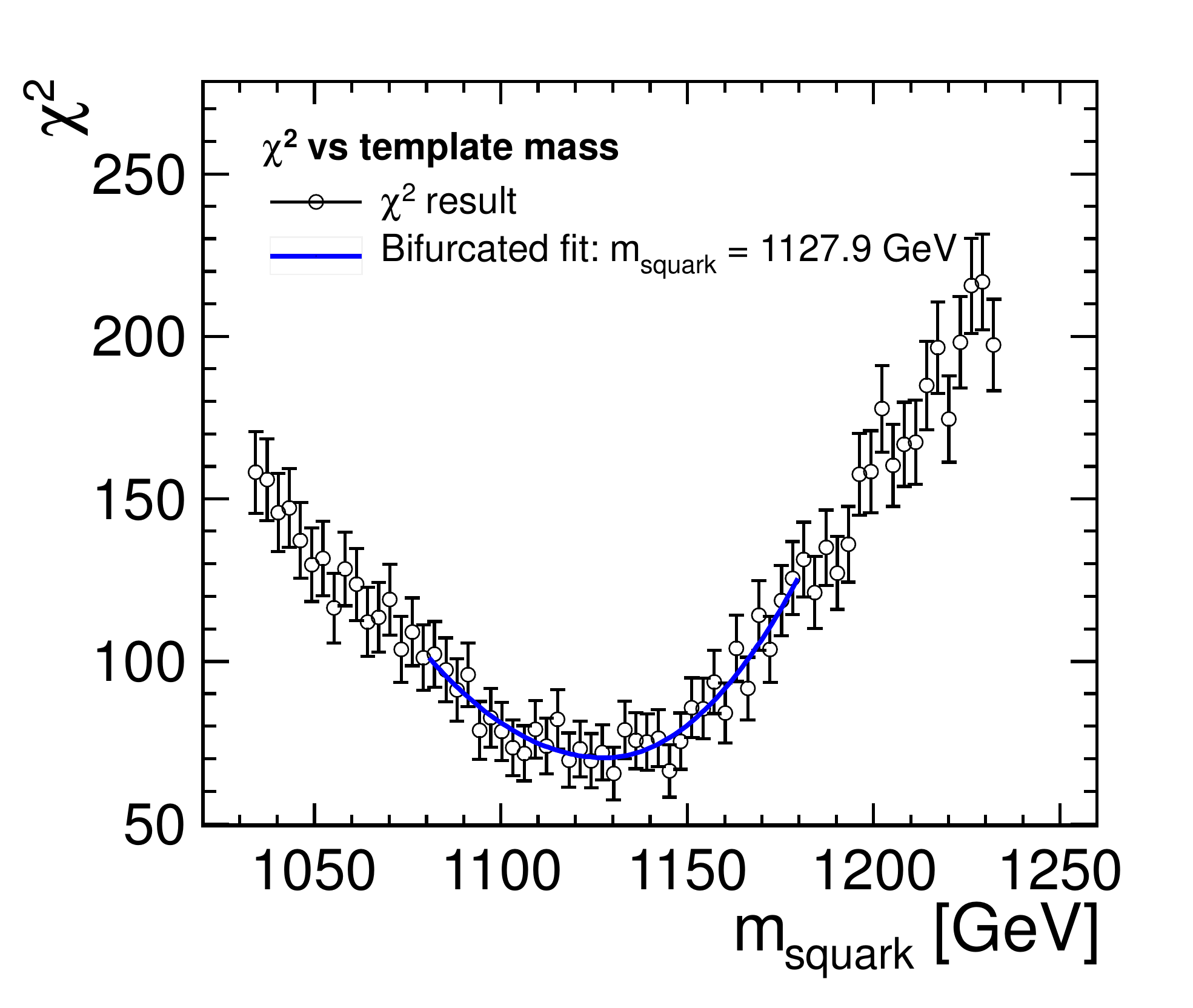}
  
  \caption{{\it Left:} The $M_C$ distribution for the measurement (background subtracted), compared to the template with the lowest $\chi^2$. {\it Right:} $\chi^2$ between
  measurement and template as a function of the squark mass in the templates. The distribution is fitted with a bifurcated parabola, with the minimum giving the measured squark mass.}
  \label{fig:TemplateFit}
\end{center}
\end{figure}

It is possible to extract the mass of the squarks using the upper edge $M_C^{max}$ of the $M_C$ distribution. For a reliable fit of the edge, the detector resolution, distortions due to the beam energy spectrum and the influence of machine-related backgrounds need to be accounted for in the fit function. Still, even small background contributions in the region of the edge can have a significant influence, resulting in biased results. 

It thus seems advantageous to use a template fit instead, which allows the inclusion of the above-mentioned effects in the generation of the templates. In such a fit, the mass is determined by comparing the observed distribution with high-statistics signal templates generated for various different squark masses. An additional advantage of a template fit is that the complete $M_C$ distribution enters into the fit, not just the high-$M_C$ edge (although the edge also is the driving region in a template fit), potentially leading to reduced statistical errors for statistically limited samples and resulting in higher stability against remaining background contributions and statistical fluctuations.

In the model used for this analysis the up-type right-squarks
($\tilde{u}_R, \tilde{c}_R$) with mass $m_{\tilde{u}}$ are about 10 GeV
heavier than their down-type counterparts ($\tilde{d}_R, \tilde{s}_R$)
with mass $m_{\tilde{d}}$.
Since the present analysis is unable to distinguish between up and down type
squarks,  the mass measurement will give the result of the mean
$m_{\rm squark}$, weighted with the respective production cross sections. Since up-type squark production has an approximately four times higher cross section than down-type squark production, the mass value is dominated by the up-type squarks.

For the template generation, a similar mass splitting between up- and down-type
squarks of exactly 10 GeV  was used. The templates were created in steps of
3 GeV ranging from 1050 GeV to 1248 GeV.

In order to minimize statistical fluctuations in the templates, each of these
mass points was generated with 50\,000 events, corresponding to an integrated luminosity
of ${\cal L} = 33.6 $ab$^{-1}$ at the true squark mass. Due to computational limitations it was not possible to perform a full
simulation and reconstruction of these 3.3 million events. Instead, detector effects were included on generator level.

As a first step, acceptance was taken into account by rejecting particles with 
$| \cos \theta | > 0.995$ or $p < 100$ MeV. Then, jet clustering was performed using  the same algorithm as used
in the rest of this analysis.
To account for detector resolution effects, the reconstructed jet energies are then smeared with a Gaussian, assuming a jet
energy resolution of 4.5\%. This resolution, as well as a small off-set in $M_C$ originating from the presence of $\gamma\gamma \rightarrow$ hadrons background which is not included in the generator-level templates, is determined by comparing the $M_C$ distribution for a template generated with the true mass values with a fully simulated high-statistics signal sample. The generated events are passed through the event selection procedure, including missing momentum requirements and the boosted decision tree, to fully reproduce the effects of the analysis procedure on the distribution.

The template fit itself is performed by comparing the $M_C$ distributions of different templates and the background-subtracted measurement using a binned $\chi^2$ with a free overall normalization of the template. The template with the lowest $\chi^2$ is shown in Figure \ref{fig:TemplateFit}\,{\it left}, compared to the data distribution. From the dependence of the $\chi^2$ on the template mass, shown in Figure \ref{fig:TemplateFit}\,{\it right}, the measured squark mass is determined through a fit with a bifurcated parabola. The statistical errors are determined with a toy MC, resulting in 
\begin{eqnarray*}
m_{\rm squark} & = & 1127.9\, \mathrm{GeV}\, \pm\, 5.9\, \mathrm{GeV}\,(stat)\, ,
\end{eqnarray*}
in good agreement with the cross-section averaged generator value of 1123.7 GeV. The statistical error for an integrated luminosity of 2 ab$^{-1}$ corresponds to 0.52\%. The error on the neutralino mass, taken from other measurements \cite{Blaising:2012vd} also enters into the present measurement. For the particle masses
considered here, a 1\,GeV uncertainty on the neutralino mass translates
into a 0.54\,GeV uncertainty on the squark mass, resulting in an error of 1.8\,GeV for the precision expected from slepton measurements with an integrated luminosity of 2 ab$^{-1}$. 

Using the selection efficiency obtained from the training phase of the boosted decision trees, the cross section was determined from the integral of the background-subtracted $M_C$ distribution.  Here, a statistical error of 0.07 fb, corresponding to 4.6\%, was achieved.

Systematic errors have not yet been evaluated thoroughly. A first study of possible systematics originating from the precision with which the beam energy spectrum can be measured has shown that, due to the use of the variable $M_C$, the effect is negligible compared to the expected statistical errors. Larger effects are expected from detector and reconstruction uncertainties, such as the jet energy scale.

\section{Conclusions}

A 3 TeV $e^+e^-$ collider based on CLIC technology allows the measurement of TeV-scale light-flavored right-squarks through the decay into a quark and the lightest neutralino, the dominant decay channel if the decay into gluinos is forbidden. Using a combination of missing energy and multivariate classifies it was possible to achieve a high signal significance despite standard-model background processes that exceed the signal production cross section by almost four orders of magnitude. The machine-related background from $\gamma\gamma \rightarrow {\rm hadrons}$ processes could be controlled by timing cuts in the reconstruction and by a suitable choice of the jet finder. For full detector simulations of the CLIC CDR SUSY benchmark model with light-flavored right-squark masses of around 1125 GeV, a statistical precision of 5.9 GeV, corresponding to 0.52\%, is achieved for combined up- and down-type squarks with an integrated luminosity of 2\,ab$^{-1}$ using a template fit with generator-level templates. For the same data sample, a statistical precision of 5\% is achieved for the total production cross section, demonstrating that precision measurements of the properties new, strongly interacting particles are possible at CLIC in a rather generic new physics signature of two energetic jets and missing energy.

\begin{footnotesize}


\end{footnotesize}


\begin{thebibliography}{99}

\bibitem{Aad:2011ib} 
  G.~Aad {\it et al.}  [ATLAS Collaboration],
  arXiv:1109.6572 [hep-ex] (2011).
  
\bibitem{Chatrchyan:2011zy} 
  S.~Chatrchyan {\it et al.}  [CMS Collaboration],
  Phys.\ Rev.\ Lett.\  {\bf 107}, 221804 (2011).



\bibitem{Assmann:2000hg}
R.~W. Assmann {\it et al.}, 
CERN-2000-008 (2000).

\bibitem{LCD-2011-016}
M.~Thomson {\it et al.}, LCD-2011-016 (2011).


\bibitem{CLIC_CDR}
L.~Linssen, A.~Miyamoto, M.~Stanitzki, H.~Weerts (ed.),Physics and Detectors at CLIC: CLIC Conceptual Design Report, {CERN-2012-003}.

\bibitem{Thomson:2009rp} 
  M.~A.~Thomson,
  Nucl.\ Instrum.\ Meth.\ A {\bf 611}, 25 (2009).

\bibitem{Cacciari:2005hq} 
  M.~Cacciari and G.~P.~Salam,
  Phys.\ Lett.\ B {\bf 641}, 57 (2006).


\bibitem{Blaising:2012vd} 
  J.~-J.~Blaising, M.~Battaglia, J.~Marshall, J.~Nardulli, M.~Thomson, A.~Sailer and E.~van der Kraaij,
  arXiv:1201.2092 [hep-ex].

\bibitem{LCD-2010-012}
F.~Simon, LCD-2010-012 (2010).


\bibitem{Tovey:2008ui} 
  D.~R.~Tovey,
  JHEP {\bf 0804}, 034 (2008).



\bibitem{Kilian:2007gr} 
  W.~Kilian, T.~Ohl and J.~Reuter,
  Eur.\ Phys.\ J.\ C {\bf 71}, 1742 (2011).

\bibitem{Sjostrand:2006za} 
  T.~Sjostrand, S.~Mrenna and P.~Z.~Skands,
  JHEP {\bf 0605}, 026 (2006).


\bibitem{Agostinelli:2002hh} 
  S.~Agostinelli {\it et al.}  [GEANT4 Collaboration],
  Nucl.\ Instrum.\ Meth.\ A {\bf 506}, 250 (2003).

\bibitem{CLIC_ILD}
A.~M\"unnich, A.~Sailer, LCD-2011-002 (2011).


\bibitem{:2010zzd} 
  T.~Abe {\it et al.}  [ILD Concept Group - Linear Collider Collaboration],
  arXiv:1006.3396 [hep-ex] (2010).

\bibitem{Hocker:2007ht} 
  A.~Hocker, J.~Stelzer, F.~Tegenfeldt, H.~Voss, K.~Voss, A.~Christov, S.~Henrot-Versille and M.~Jachowski {\it et al.},
  PoS ACAT {\bf }, 040 (2007).



\end{thebibliography}
\end{document}